\documentclass[prx,letterpaper,nobalancelastpage,twocolumn,superscriptaddress,nofootinbib]{revtex4-1}
\usepackage{amsmath,amssymb}
\usepackage{graphicx}
\usepackage{color} 
\definecolor{LinkColor}{rgb}{0,0,.5}
\usepackage[colorlinks=true,citecolor=LinkColor,urlcolor=LinkColor]{hyperref}

\newcommand{\ket}[1]{|#1\rangle}
\newcommand{\bra}[1]{\langle #1 |}
\newcommand{\tr}[1]{\text{tr}\left(#1\right)}

\graphicspath{{./figures/}}

\begin{document}
\title{Environment-assisted quantum-enhanced sensing with electronic spins in diamond}

\author{Alexandre Cooper}
\affiliation{Department of Nuclear Science and Engineering and Research Lab of Electronics,\\Massachusetts Institute of Technology, Cambridge, MA 02139, USA }
\affiliation{Department of Physics, Mathematics and Astronomy,\\California Institute of Technology, Pasadena, CA 91125, USA }

\author{Won Kyu Calvin Sun}
\author{Jean-Christophe Jaskula}
\author{Paola Cappellaro}\email{pcappell@mit.edu}
\affiliation{Department of Nuclear Science and Engineering  and Research Lab of Electronics,\\Massachusetts Institute of Technology, Cambridge, MA 02139, USA }

\date{\today}

\begin{abstract}
The performance of solid-state quantum sensors based on electronic spin defects is often limited by the presence of environmental spin impurities that cause decoherence. A promising approach to improve these quantum sensors is to convert environment spins   into useful resources for sensing. Here we demonstrate the efficient use of an unknown electronic spin defect in the proximity of a nitrogen-vacancy center in diamond as both a quantum sensor and a quantum memory.  We first 
experimentally evaluate the improvement in magnetic field sensing provided by mixed entangled states of the two electronic spins. Our results critically highlight the tradeoff between the advantages expected from increasing the number of spin sensors and the typical challenges associated with increasing control errors, decoherence rates, and time overheads. Still, by taking advantage of the spin defect as both a quantum sensor and a quantum memory whose state can be repetitively measured to improve the readout fidelity, we can achieve a  gain in performance  over the use of a single-spin sensor. 
These results show that the efficient use of available quantum resources can  enhance quantum devices, pointing to a practical strategy towards quantum-enhanced sensing and  information processing by exploiting environment spin defects.
\end{abstract} 
\maketitle

\section{Introduction}
Precision measurement of weak magnetic fields at the atomic scale using spin defects in solids is enabling novel applications in the physical and life sciences~\cite{Taylor2008, Schirhagl2014, Degen2017}.
Nitrogen-vacancy (NV) centers in diamond are particularly suitable for sensing magnetic fields, as their electronic spins can be optically polarized and read out, as well as coherently controlled under ambient conditions over long coherence times.
Such spin sensors have recently been used for characterizing magnetic thin films~\cite{Maletinsky2012, Gross2017a}, imaging living cells~\cite{LeSage2013}, detecting single molecules~\cite{Lovchinsky2016}, and performing nuclear magnetic resonance spectroscopy of small-volume chemical samples~\cite{Mamin2013, Aslam2017, Glenn2018}.

An important challenge in increasing the sensitivity of spin sensors is taking advantage of their intrinsic quantum nature~\cite{Tilma2010, Giovannetti2011}, e.g., to realize spin-squeezed states~\cite{Wineland1992, Kitagawa1993,Esteve2008} or entangled states~\cite{Jones2009} of $n$ spins, which improve the precision by $\sqrt{n}$ over the use of $n$ independent spins~\cite{Bollinger1996, Huelga1997}.
This requires ensembles of interacting spins that  can be efficiently initialized, manipulated, and read out, as well as robustly prepared in entangled states with long coherence times.
Quantum-enhanced sensing with spin defects in diamond has been so far prevented by the difficulty of accessing ensembles of strongly-coupled NV centers~\cite{Neumann2010, Dolde2013}, while mitigating the detrimental influence of nearby defects on their coherence and charge state properties.

\begin{figure}[t!]
\centering
\includegraphics[width=\columnwidth]{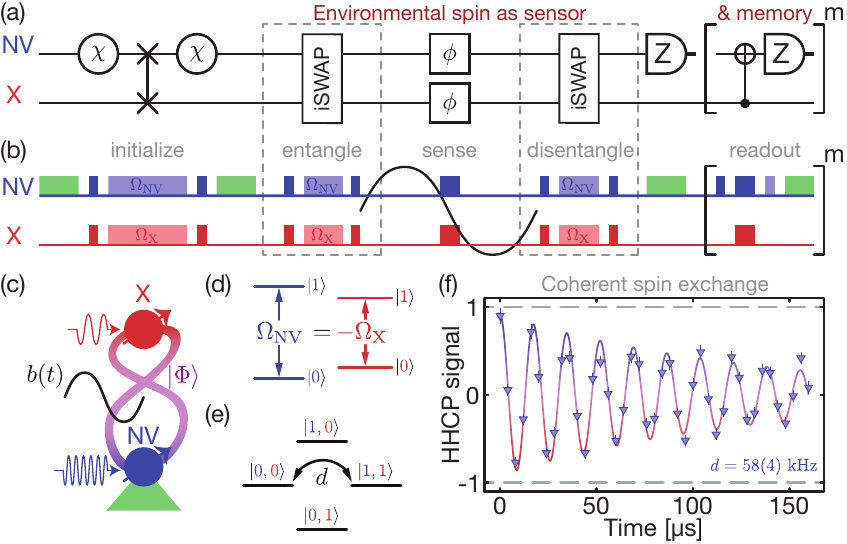}
\caption{
\textbf{Environment-assisted quantum-enhanced sensing.}
{(a-c)}~Circuit, pulse sequence, and experimental system for sensing magnetic fields with a mixed entangled state of two electronic spins associated with a nitrogen-vacancy (NV) center and an electron-nuclear spin defect (X) in diamond. The two spins are polarized with a dissipative channel $\chi$ acting on the NV spin (green laser pulse) and a coherent spin-exchange gate between the NV and X spins (cross-polarization sequence). A mixed entangled state $\rho_\Phi$, 
prepared with an entangling gate (cross-polarization sequence), acquires a coherent phase $\phi_\Phi=\phi_\text{NV}+\phi_\text{X}=2\phi$, upon interacting with a sinusoidal magnetic field $b(t)$, which is then mapped as a population difference onto both the NV and X spins (cross-polarization sequence) and read out by performing a projective measurement on the NV spin (green laser pulse) and a series of $m$ repetitive measurements on the X spin (recoupled spin-echo sequence). Resonant microwave pulses selectively drive the NV and X spins, whereas green laser pulses excite the NV spin for polarization and readout, while leaving the X spin unaffected.
{(d-f)}~The spectrally mismatched spins are continuously driven at the Hartmann-Hahn matching condition, $|\Omega_{\text{NV}}|=|\Omega_{\text{X}}|$, to induce coherent spin exchange at the dipolar coupling strength $d=58(4)~\text{kHz}$ with a decay time of $T_{1\rho}=132(11)~\mu\text{s}$.
}
\label{FigExperimentalSystem}
\end{figure}

To overcome these challenges, we explore an approach to quantum-enhanced sensing based on exploiting electronic spins in the environment of a single NV spin-sensor~\cite{Schaffry2011, Goldstein2011}, such as those associated with crystalline defects~\cite{Shi2013, Grinolds2014, Knowles2016, Rosenfeld2018, Cooper2018a}, surface spins~\cite{Sushkov2014, Sangtawesin2018}, or paramagnetic labels~\cite{Shi2015,Schlipf2017}.
Besides increasing the number of sensing spins, combining diverse spin species enables distributing sensing tasks, e.g., the primary spin is used for state preparation and read out, while auxiliary spins are used for sensing and storage.
This approach bears resemblance to mixed-species quantum logic with trapped atomic ions~\cite{Schmidt2005, Ballance2015}, where an auxiliary ion is used to cool, prepare, and read out the state of another ion, which in turn serves the role of a memory or a spectroscopic probe.

To demonstrate our approach to environment-assisted quantum-enhanced sensing, we focus on the problem of measuring time-varying magnetic fields with mixed entangled states~\cite{Vedral1997} of two electronic spins~(Fig.~\ref{FigExperimentalSystem}a). Specifically, we perform magnetic sensing experiments with a mixed entangled state of two electronic spins associated with a single NV center and an electron-nuclear spin defect (X) in diamond~(Fig.~\ref{FigExperimentalSystem}b-c).
The X defect is one of two environmental spin defects, with electronic spin $S=1/2$ and nuclear spin $I=1/2$, whose hyperfine and dipolar interaction tensors have been recently characterized~\cite{Cooper2018a}. Taking advantage of its stability under optical illumination, we exploit the X spin as both a sensor and a memory whose state can be repetitively measured to improve readout fidelity.

\section{Quantum control} 
To illustrate our ability to convert environmental spin defects into resources for sensing, we first implement coherent control techniques to initialize, control, and read out the X spin via the NV spin, as well as generate and characterize a mixed entangled state of two spins, which are known to contribute to quantum enhancement in metrology~\cite{Modi2011}.

All experiments are performed at room temperature with a static magnetic field of $205.2(1)~\text{G}$ aligned along the molecular axis of the NV center. Because of the large energy mismatch between the NV and X spins~(Fig.~\ref{FigExperimentalSystem}d), coherent spin exchange in the laboratory frame is suppressed. We optically polarize and read out the NV spin using green laser pulses; because the X spin lacks a physical mechanism for state preparation and detection, while being robust against optical illumination, we use the NV spin to initialize and repetitively read out its state using cross-polarization and recoupled spin-echo sequences~(Fig.~\ref{FigExperimentalSystem}b).
Both NV and X spins are coherently controlled using resonant microwave pulses delivered through a coplanar waveguide.
We address only one out of two hyperfine transitions of the X spin to reduce control errors and time overheads.
Because the X nuclear spin is unpolarized, our nominal signal contrast is reduced by half; throughout the manuscript, we normalize our signal by subtracting its nominal baseline and multiplying it by a factor of two, but analyze the performance of our approach for both an unpolarized and a fully polarized X nuclear spin.

\begin{figure}[t!]
\centering
\includegraphics[width=\columnwidth]{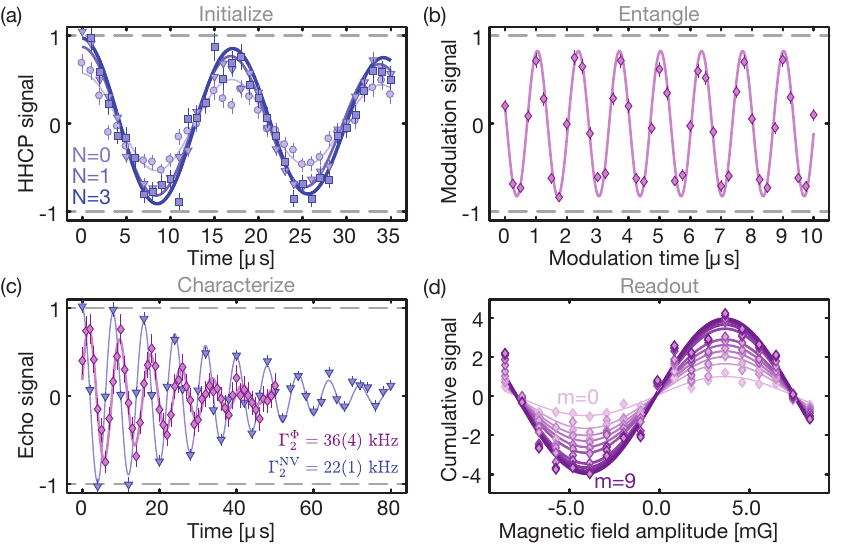}
\caption{
\textbf{Quantum control.}
{(a)}~The increase in contrast of the cross-polarization signal after $N=0$ (circle), $N=1$ (triangle), and $N=3$ (square) rounds of polarization transfer indicates an increase in X spin polarization from $14(3)\%$ to $76(3)\%$ and $94(6)\%$ respectively.
{(b)}~The coherent oscillations of the modulation signal measured after applying the entangling and disentangling gates indicate the preparation and detection of two-spin coherence 
with a contrast of $85(3)\%$. The phases of the pulses of the disentangling gate are modulated at $500~\text{kHz}$ and $250~\text{kHz}$ for the NV and X spins, such that the signal oscillates at $751(2)~\text{kHz}$, the sum of both frequencies.
{(c)}~The decay of the spin-echo signal (purple diamonds) for the mixed entangled state corresponds to a two-spin decoherence rate of $\Gamma_2^{\Phi} = 36(4)$~kHz, which is consistent with the sum of the decoherence rates for the NV spin (blue triangles, $\Gamma_2^{\text{NV}} = 22(1)$~kHz) and X spin (not shown, $\Gamma_2^{\text{X}} = 15(2)$~kHz).
{(d)}~Cumulative magnetometry signal measured for a sensing time of $\tau=19~\mu\text{s}$ using $m=0$ (light purple diamonds) to $m=9$ (dark purple diamonds) repetitive measurements of the X spin. The cumulative signal is normalized to the maximum amplitude of the magnetometry signal measured without repetitive measurements ($m=0$). 
The cumulative signal contrast increases by a factor of $4.2$ after $m=9$ repetitive measurements, providing an increase in signal-to-noise ratio of 1.91(8) using optimal weights.
}
\label{FigQuantumControl}
\end{figure}
 
\subsection{State initialization via polarization transfer}

We initialize the X spin by transferring polarization from the NV spin using a Hartmann-Hahn Cross-Polarization (HHCP) sequence~\cite{Hartmann1962, Laraoui2013}, during which both NV and X spins are continuously driven at the same Rabi frequency, $\Omega_\text{NV}=-\Omega_\text{X}$, such that both spins are brought into resonance in the rotating frame~(Fig.~\ref{FigExperimentalSystem}e). The cross-polarization sequence introduces coherent spin exchange between the NV and X spins at the dipolar coupling strength $d=58(4)~\text{kHz}$~(Fig.~\ref{FigExperimentalSystem}f), which we use with an appropriate choice of driving phases to implement both polarization transfer gates (SWAP, $|00\rangle\mapsto|11\rangle$) and entangling gates (iSWAP, $|00\rangle\mapsto(|00\rangle\pm i|11\rangle)/\sqrt{2}$) to prepare and detect two-spin coherence for sensing~\cite{Schuch2003}. Using the HHCP sequence for a spin-exchange time of $\tau_\text{HHCP}=8.6~\mu\text{s}$ after a green laser pulse, we perform $N$ rounds of polarization transfer~(Fig.~\ref{FigQuantumControl}a). The contrast of the cross-polarization signal increases from $49(3)\%~(\text{N}=0)$ to $82(3)\%~(\text{N}=1)$ and $88(4)\%~(\text{N}=3)$, corresponding to an increase in X spin polarization from $14(3)\%$ to $76(3)\%$ and $94(6)\%$ respectively.
Because of the tradeoff between increasing polarization and reducing time overheads, we perform all of our experiments after a single round of polarization transfer.

\subsection{Entanglement generation and characterization}
To generate the mixed entangled state $\rho_\Phi$, we then implement an entangling gate using the HHCP sequence for half the spin-exchange time of $\tau_\text{HHCP}/2=4.3~\mu\text{s}$~(Fig.~\ref{FigQuantumControl}b).
While we cannot prepare the pure Bell entangled states, $\ket{\Phi_\pm}=(\ket{00}\pm i\ket{11})/\sqrt{2}$, 
we can still achieve a mixed state that has  non-zero two-spin coherence in the subspace spanned by the Bell states, i.e., $\tr{\rho_\Phi\ket{\Phi_\pm}\bra{\Phi_\pm}}\neq0$.
Though such mixed states are unavoidable due to experimental imperfections, they still prove to be useful resources for sensing applications, as we demonstrate below.

We characterize the two-spin coherence by converting it back into a population difference of the NV spin using a modulated disentangling gate; we modulate the phases of the pulses of the cross-polarization sequence acting on the NV and X spins at $500~\text{kHz}$ and $250~\text{kHz}$ to impart coherent oscillations at the sum of both frequencies~\cite{Mehring2003a, Scherer2008}, simulating the evolution of two-spin coherence in the presence of a static magnetic field. As expected, the signal oscillates at $751(2)~\text{kHz}$, the sum of both frequencies, and the signal contrast is $85(3)\%$, consistent with the value measured after a single round of polarization transfer~(Fig.~\ref{FigQuantumControl}b).

We further measure the coherence time of the single-spin and two-spin states~(Fig.~\ref{FigQuantumControl}c) using a spin-echo sequence with decoupling pulses applied simultaneously to both NV and X spins.
The coherence signal is fitted to $S(\tau)\propto e^{-(\Gamma_2\cdot \tau)^p}$, where $\Gamma_2$ is the decoherence rate and $p\approx1.6$ is the decay exponent.
We measure 
$\Gamma_2^\text{NV}=22(1)~\text{kHz}$, $\Gamma_2^\text{X}=15(2)~\text{kHz}$, and $\Gamma_2^{\Phi} = 36(4)~\text{kHz}$, such that the decoherence rate of the two-spin state is consistent with the sum of the decoherence rates of the NV and X spins, $\Gamma_2^{\Phi}=\Gamma_2^{\text{NV}}+\Gamma_2^{\text{X}}$.
The NV and X spins experience different magnetic environments, such that $\Gamma_2^\text{X}<\Gamma_2^\text{NV}$, which is beneficial for achieving a gain in sensitivity using environmental spin defects.

\subsection{Repetitive readout}
We finally demonstrate repetitive readout of the X spin via the NV spin. Any arbitrary state of the X spin can be generally measured by coherently mapping it onto the state of the NV spin before optical readout, e.g., using a cross-polarization sequence, as done for the experiments in Fig.~\ref{FigQuantumControl}a,c. 
A population state of the X spin can also be measured by correlating the states of the NV and X spin before optical readout, e.g., using a recoupled spin-echo sequence~(Fig.~\ref{FigExperimentalSystem}b). 

Conversely, we can improve the  NV spin  readout by exploiting the X spin as a quantum memory, storing the desired information onto its state:  
as the X spin is stable under optical illumination, we can repeat multiple times the mapping and readout steps  to increase the signal-to-noise ratio~\cite{Jiang2009}. 
For example, the amplitude of the optimally-weighted cumulative signal obtained for a magnetometry experiment at $\tau=19~\mu\text{s}$ increases by a factor of 4.2 after $m=9$ additional repetitive measurements, providing an improvement in signal-to-noise ratio of 1.91(8)~(Fig.~\ref{FigQuantumControl}d). We note that, in the current experiment, the number of repetitive measurements that provides an improvement in signal-to-noise ratio is not limited by the intrinsic relaxation of the X spin, but by imperfections in the gate used for the mapping.
These results illustrate the advantage of working with environmental spins that are robust against optical illumination, such that they can be used not only as quantum sensors, but also as quantum memories that can be repetitively measured. 
 
\section{Magnetic sensing}
We now focus on the problem of estimating the amplitude of a time-varying magnetic field whose temporal profile is known, here a sinusoidal field $b(t)=b\sin{(2\pi\nu t)}$.
We sample the field with a phase-matched spin-echo sequence of duration $\tau=1/\nu$ with decoupling pulses applied simultaneously on both spins.
The average signal expected for $n$ maximally entangled spins, which we assume are all equally coupled to the field, is given by $S_n(\tau)\propto\alpha_n(\tau)\sin(\nu_n(\tau)b)$, where $\nu_n(\tau)=n\gamma_e\hat{f}\tau$ is the precession rate of the interferometric signal, $\gamma_e=2\pi\cdot2.8~\text{MHz}$ is the gyromagnetic ratio of the electronic spin, and $\hat{f}\leq2/\pi$ is a scaling factor quantifying the overlap between the sinusoidal field and the spin-echo sequence, with the equality held when phase-matched~\cite{Taylor2008, deLange2011, Magesan2013, Cooper2014}.

\begin{figure}[t!]
\centering
\includegraphics[width=\columnwidth]{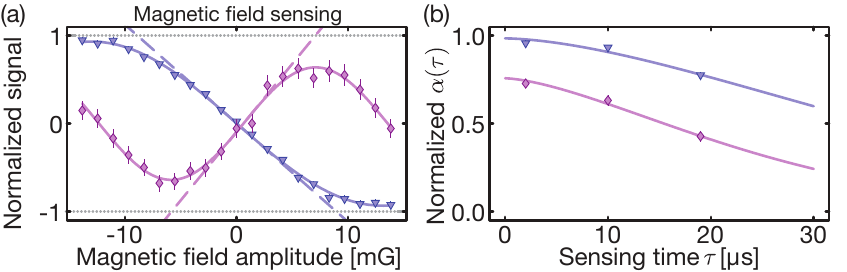}
\caption{
\textbf{Magnetic sensing.}
{(a)}~Magnetometry signal measured with a single NV spin (blue triangles) and a mixed entangled state of two spins (purple diamonds) obtained by varying the amplitude of a sinusoidal magnetic field for a sensing time of $\tau=10~\mu\text{s}$.
The two-spin state precesses at twice the rate as the single-spin state.
The sensitivity is proportional to the slope (dash lines) of the sinusoidal fit (solid curves). The magnetometry signal for the NV spin has been inverted for clarity.
{(b)}~The decrease in contrast of the magnetometry signal (normalized amplitude) for the single-spin state (blue triangles) and the two-spin state (purple diamonds) measured for a sensing time of $\tau=2$, $10$, and $19~\mu\text{s}$ is explained by decoherence during sensing (solid curves, which are fits to the echo decays in Fig.~\ref{FigQuantumControl}). The nominal loss in amplitude, $\alpha^{\text{NV}}_0=96(3)\%$ and $\alpha_0^{\Phi}=78(6)\%$, is instead explained by dissipation and unitary control errors.
}
\label{FigMagneticSensing}
\end{figure}

We measure the magnetometry signal by sweeping the amplitude of the sinusoidal field~(Fig.~\ref{FigMagneticSensing}a) for a sensing time of $\tau=2$, $10$, and $19~\mu\text{s}$.
The signal shows coherent oscillations with the two-spin state precessing at twice the rate of the single-spin state.
The relative difference in signal contrast is explained by decoherence during sensing and control errors~(Fig.~\ref{FigMagneticSensing}b). Indeed, the decrease in signal amplitude satisfies $\alpha_n(\tau) = \alpha_0^n e^{-(\Gamma_2^n\cdot \tau)^{p_n}}$, where $\Gamma_2^n$ and $p_n$ are the coherence parameters for $n$-spin states measured from independent spin-echo experiments~(Fig.~\ref{FigQuantumControl}c).
We estimate a nominal amplitude of $\alpha_0^\text{NV}=96(3)\%$ and $\alpha_0^{\Phi}=78(6)\%$ for the single-spin and two-spin states, resulting from inefficiencies during state preparation and readout caused by dissipation and unitary control errors.

\section{Performance analysis}
We finally quantify the improved performance of the composite quantum sensor with respect to a single spin sensor,  
analyzing the situation in which the spin defect acts as just a sensor or both as a sensor and a memory. 

The relevant figure of merit is the smallest magnetic field $\delta b_n=\sigma_{S_n}/|dS_n(\tau)/db|$ that can be measured by the $n$-spin sensor, where  $\sigma_{S_n}$ is the standard deviation of the magnetometry signal, and $dS_n(\tau)/db=\alpha_n(\tau)\nu_n(\tau)$ is the maximum slope of the magnetometry signal~(Fig.~\ref{FigMagneticSensing}a). 
We thus define the \textit{gain in performance} as $g_n(\tau)=\delta b_1(\tau)/\delta b_n(\tau)$, which exceeds unity when the $n$-spin state outperforms the single-spin state, and is upper bounded by $n$ if utilizing $(n-1)$ ancillas as sensors only.
In our experiment, the gain in performance is given by $g_\Phi(\tau)=|\alpha_\Phi(\tau)\nu_\Phi(\tau)|/|\alpha_\text{NV}(\tau)\nu_\text{NV}(\tau)|\leq2$, as we can safely assume that the signal uncertainty $\sigma_S$ is the same for the single-spin and two-spin sensing cases, as for both cases the signal is obtained by measuring the same NV and is limited by shot noise. 

This figure of merit is relevant in many scenarios, such as when the experiment can only be repeated a fixed number of times due to external constraints, e.g. the duration or triggering of the signal to be measured. 
However, often a more practical metric is the smallest magnetic field that can be measured in a fixed time. We thus also consider the  sensitivity $\eta=\delta b\sqrt{T}$, where $T=M(\tau+\tau_O)$  is the total experimental time for $M$ measurements with  a sensing time  $\tau$  and $\tau_O$  time overheads. 
To account for these  time overheads, needed to prepare and readout the mixed entangled state of $n$ spins, we define the \textit{gain in sensitivity} as $\tilde{g}_n(\tau)=g_n(\tau)h_n(\tau)$, where $g_n(\tau)$ is the gain in performance and $h_n(\tau)$ is the relative time overheads for $n$-spin protocols with more complex control requirements.
In our experiment, $h_\Phi(\tau)=\sqrt{(\tau+\tau_{NV})/(\tau+\tau_\text{NV}+\tau_\Phi)}$, where $\tau$ is the sensing time, $\tau_\text{NV}=5.7~\mu\text{s}$, and $\tau_\Phi=21~\mu\text{s}$, which includes the additional time needed for state initialization, entanglement generation, and  state readout, and scales inversely with the dipolar coupling strength. 
\begin{figure}[t!]
\centering
\includegraphics[width=\columnwidth]{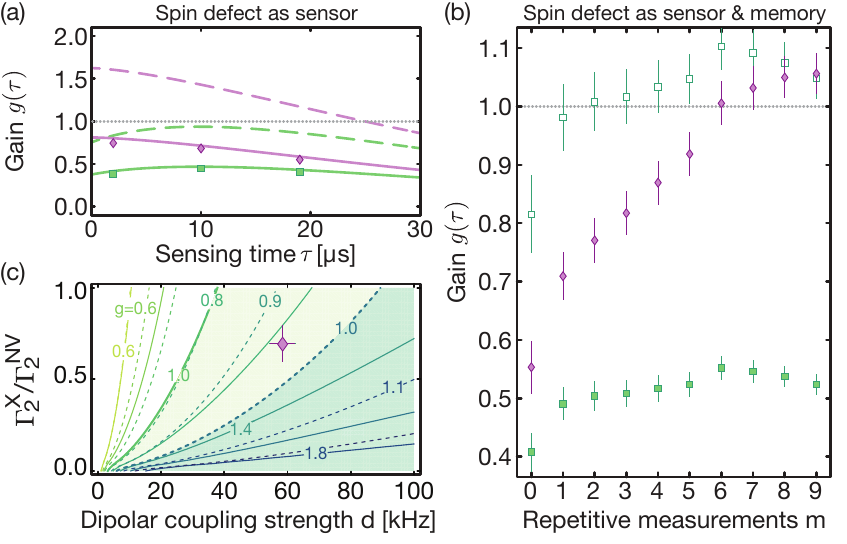}
\caption{
\textbf{Performance analysis.}
{(a)}~Measured gain in performance and sensitivity achieved using a mixed entangled state before (purple diamonds) and after (green squares) accounting for time overheads as a function of  sensing time.
The expected gain in performance for a fully polarized X nuclear spin (dashed purple line) is greater than unity for up to $\approx25~\mu\text{s}$, but is less than unity for all times when accounting for time overheads (dashed green line).
{(b)}~Using the X spin as a memory for repetitive measurements improves the performance. The measured gain in performance (purple solid diamonds) before accounting for time overheads surpasses unity after $m=6$ repetitive measurements and reaches a maximum of $g=1.06(4)$ after $m=9$ repetitive measurements. The measured gain in sensitivity maximizes to $\tilde{g}=0.55(2)$ when accounting for time overheads (green solid squares); however, the extrapolated gain in sensitivity for a fully-polarized X nuclear spin achieves a maximum gain of $\tilde{g}=1.10$ after $m=6$ repetitive measurements.
{(c)}~Maximum gain in sensitivity achievable (dashed contour lines) for different values of the dipolar coupling strengths and relative decay rates of the two-spin system used for sensing.
We assume a fully-polarized X nuclear spin, but use the control errors and NV coherence time measured experimentally.
Without repetitive measurements, our two-spin system (purple diamond) lies outside the region where it outperforms a single spin (gain in performance $g>1$, dark green region).
With repetitive measurements (solid contour curves), the increase in signal-to-noise ratio, even after accounting for additional time overheads, shifts the boundary of the region of favorable performance, $g>1$ (light green region), such that our two-spin system lies within it.
}
\label{FigPerformance}
\end{figure}

\subsection{Environmental spin as sensor}

Our results reported in Fig.~\ref{FigPerformance}a show that, despite the twofold increase in precession rate, $|\nu_\Phi(\tau)/\nu_\text{NV}(\tau)|=2$, the relative amplitude of the magnetometry signal is less than half, $|\alpha_\Phi(\tau)/\alpha_\text{NV}(\tau)|\leq1/2$, such that no gain in performance is achieved for an unpolarized X nuclear spin. However, extrapolating our data to a fully polarized X nuclear spin or to driving both hyperfine transitions simultaneously, we predict a gain in performance greater than unity for up to $\tau\approx25~\mu\text{s}$. Still, when accounting for time overheads, there exists no sensing time for which a gain in sensitivity is achievable, unless our control imperfections were reduced by at least $8~\%$.

These results illustrate the fundamental tradeoff between increasing the number of spins and increasing control errors, decoherence rates, and time overheads. 
One approach to achieving a gain in sensitivity would be to improve the control fidelity or to look for a system with more favorable parameters. For instance, our simulation results illustrated in Fig.~\ref{FigPerformance}c show that, assuming the same level of control errors, a spin defect with either stronger dipolar coupling $d\gtrsim 75$~kHz (reducing time overhead) or smaller decoherence rate, $\Gamma_2^\text{X}/\Gamma_2^\text{NV}\lesssim 0.4$ (increasing visibility at given $\tau$), is sufficient to reach the regime where the entangled state  outperforms the single spin sensor.

\subsection{Environmental spin as sensor and memory}
Here we demonstrate a different approach for improving the gain in performance, 
which exploits the X spin as both a sensor and a memory that can store information about the measured field. The stored information can be then repetitively measured with a quantum non-demolition measurement~\cite{Braginsky1980}. This results in an increased signal-to-noise ratio SNR($m$) after $m$ repetitive measurements and a upper bound for the gain in performance of $n\,\text{SNR}(m)$. In our case, the quantum non-demolition measurement is enabled by the fact that even at low magnetic field the X spin is unperturbed by the optical pulse used to perform a projective measurement on the NV spin, providing an advantage over other ancillary spin systems such as nitrogen nuclear spins or nitrogen substitutional impurities (P1 centers). In addition, going beyond the typical repetitive readout scheme~\cite{Jiang2009,Neumann2010b}, we take advantage of the fact that our disentangling gate maps the magnetic field-dependent phase as a population difference on both the NV and X spins. This provides a significant advantage, as it bypasses the need for  an 
additional mapping operation, reducing both time overheads and the loss in signal contrast caused by control imperfections.


Using repetitive measurements of the X spins, we experimentally observe a two-fold maximum increase in $\text{SNR}$ at both $\tau=2~\text{and}~19~\mu\text{s}$, such that a gain in performance is achieved for the entire coherence time of the two-spin sensor. For sensing experiments at $\tau=19~\mu\text{s}$ (Fig.~\ref{FigPerformance}b), we achieve a gain in performance greater than unity after $m=6$ repetitive measurements with a maximum of $g_\text{rr}=1.06(4)$ after $m=9$ repetitive measurements. 
When accounting for (increased) time overheads, using $\tau_\Phi\mapsto\tau_\Phi+(m-1)\tau_\text{rr}$ with $\tau_\text{rr}=6.1~\mu\text{s}$, the gain in sensitivity maximizes to $\tilde{g}_\text{rr}=0.55(2)$ after $m=7$ measurements when driving only one hyperfine transition of the unpolarized X nuclear spin. We expect to reach $\tilde{g}_\text{rr}=1.10(4)$ for a fully polarized X nuclear spin or by driving both transitions. Further gains in sensitivity could be achieved by accessing more strongly coupled spin systems with longer coherence times (see Fig.~\ref{FigPerformance}c), and reducing control errors, e.g., using optimal control techniques~\cite{Dolde2014}. 


\section{Conclusions}
In conclusion, we have experimentally demonstrated an approach to quantum-enhanced sensing using mixed entangled states of electronic spins by converting electron-nuclear spin defects in the environment of a single-spin sensor into useful resources for sensing, serving both as quantum sensors and quantum memories whose state can be repetitively measured. This approach complements ongoing efforts to improve the performance of single-spin sensors, including methods to  limit the concentration of spin impurities~\cite{Balasubramanian2009,Kucsko18}, extend coherence times with quantum error correction~\cite{deLange2010, Knowles2014, Kessler2014,Bar-Gill2013}, and increase collection efficiency with photonic structures~\cite{Babinec2010,Wan18}. We emphasize that this approach is not specific to our electron-nuclear spin defect, but also applicable to other environmental spin defects with favorable coupling strengths, coherence times, and stability under optical illumination. 

Using coherent control protocols to initialize and repetitively read out a single-spin defect, as well as create a mixed entangled state with two-spin coherence, 
we achieve a gain in performance in sensing time-varying magnetic fields and predict a gain in sensitivity for a fully polarized X nuclear spin, which is within experimental reach by further extending our control toolbox to nuclear spins. 
Still, we find that common challenges associated with increased control errors, faster decoherence of entangled states, and time overheads associated with their creation limit the sensitivity improvement. In particular, the additional time required for initialization, control and readout of the entangled state is especially deleterious, a practical fact often overlooked that our study helps highlighting.
To at least partially overcome these challenges, we demonstrate that the environmental spin defect can serve a dual role, not only acting as a magnetic field sensor, but also as a quantum memory, enabling repetitive readouts of the relative population of its spin states.

Our results thus demonstrate that, despite the increased complexity and  fragility, quantum control protocols can turn electronic spin defects in the environment of a single-spin sensor, usually considered as noise sources, into useful resources for realizing quantum-enhanced sensing. 
Further improvements in quantum control, such as optimal control techniques to improve gate fidelities~\cite{Dolde2014}, and materials properties, e.g.,  to deterministically create confined ensembles of interacting spin defects with slower decoherence rates and stronger coupling strengths, should enable achieving magnetic sensing beyond the standard quantum limit using electronic spin sensors~\cite{Jones2009}.

\section{Acknowledgements}
This work was in part supported by NSF grants PHY1415345 and EECS1702716.
A.~C. acknowledges financial support by the Fulbright Program and the Natural Sciences and Engineering Research Council of Canada.
We are grateful to Chinmay Belthangady and Huiliang Zhang for their experimental support.

\bibliographystyle{apsrev4-1}
\bibliography{biblioMendeley}

\end{document}